# GATECHAIN: A BLOCKCHAIN-BASED APPLICATION FOR COUNTRY ENTRY–EXIT REGISTRY MANAGEMENT

**Mohamad AKKAD**
Düzce University, Institute of Graduate Studies, Computer Engineering Master's with Thesis
ORCID: 0009-0004-8726-2763

**Dr. Öğr. Üyesi Hüseyin BODUR**
Düzce University, Faculty of Engineering, Department of Computer Engineering
ORCID: 0000-0002-2815-3397

## ABSTRACT

Recording entry and exit records for a country, with properties such as confidentiality, integrity, and auditability, is increasingly important due to rising international mobility and security requirements. Traditional border control systems, which rely on centralised databases, are vulnerable to data manipulation and have limited interoperability between institutions. This study presents GateChain, a blockchain-based application that addresses these vulnerabilities. GateChain aims to enhance data integrity, reliability, and transparency by recording entry and exit events on a distributed, immutable, and cryptographically verifiable ledger. The application provides real-time access control and verification for authorised institutions. This paper describes the architecture and security components of GateChain and evaluates its performance and security features.

## INTRODUCTION

During entry and exit procedures, information such as nationality, name, passport number, and entry or exit date is recorded in a confidential, secure, and data-integrity database. Traditional border control systems typically store sensitive travel data in centralised databases managed and operated by government agencies. These systems are vulnerable to unauthorised data access, limited interoperability between agencies, and potential data loss. Furthermore, real-time verification requirements and the increasing volume of global travel complicate data integrity, scalability, and auditability in centralised systems.

Blockchain technology, applied in sectors such as healthcare, transportation, logistics, smart grids, smart cities, agriculture, real estate, e-voting, and cryptocurrency transactions, provides immutable, decentralised, and cryptographically verifiable data structures.

By eliminating reliance on a single trusted authority, the distributed ledger structure enables secure, transparent, and tamper-evident record management across multi-stakeholder systems. For entry and exit security, blockchain ensures records are tamper-resistant, easily auditable, and secure. It also enables controlled data sharing among national institutions and authorised third parties. These features make blockchain a strong candidate for next-generation border management infrastructures.

This study proposes a blockchain-based application, GateChain, to enhance the integrity and operational efficiency of entry and exit records. GateChain uses a distributed ledger to securely and immutably record and verify individuals' border crossing events. The system provides real-time access control and verification mechanisms for authorised institutions and supports interoperability with existing border management systems. Through cryptographic hash matching and immutable record-keeping, GateChain aims to minimise data manipulation due to fraud and increase the traceability of border crossings.

## LITERATURE VIEW

Blockchain technology is used in many areas beyond cryptocurrency transactions. In one study, Javaid and colleagues examined the significance of blockchain technology for the financial sector. They emphasised that blockchain is more reliable and less costly than traditional methods for credit reporting and issuing digital securities (Javaid, Haleem, Singh, Suman and Khan, 2022). In another study, Ashfaq et al. proposed a mechanism that combines machine learning and blockchain technology for fraud detection (Ashfaq, Khalid, Yahaya, Aslam, Azar, Alsafari and Hameed, 2022). Berenjestanaki et al. reviewed security and privacy issues, as well as the applicability of blockchain-based e-voting systems (Hajian Berenjestanaki, Barzegar, El Ioini and Pahl, 2023). In their study, Das et al. proposed an integrated blockchain-based framework to address document management problems in the construction industry (Das, Tao, Liu and Cheng, 2022).

Gadekallu et al. conducted a comprehensive study on blockchain applications for the metaverse, addressing the technical challenges of the metaverse and highlighted blockchain solutions to these challenges (Gadekallu, Huynh-The, Wang, Yenduri, Ranaweera, Pham and Liyanage, 2022). Bodur and Al Yaseen proposed a blockchain-based hybrid architecture for the secure storage, access, and sharing of medical records (Bodur and Al Yaseen, 2024; Bodur and Al Yaseen, 2024, April). Ghayvat et al. proposed a blockchain-based secure key management and encryption scheme to address the privacy and authentication of healthcare stakeholder identities (Ghayvat, Pandya, Bhattacharya, Zuhair, Rashid, Hakak and Dev, 2021).

AlBataineh explored the use of smart contracts on the blockchain to enhance sustainable development in Jordan, highlighting that blockchain has the potential to address multiple land administration challenges simultaneously (AlBataineh, 2025). In another study, Priyadarshi and Sree developed a blockchain-based system comprising three independent modules to address the digitisation, security, and storage issues of the land registry system in India (Priyadarshi and Sree, 2025). In their study, Wang et al. proposed an architecture for agricultural applications that combines blockchain with artificial intelligence, remote sensing, and the Internet of Things (IoT) to increase sustainability and productivity (Wang, Wu, Zeng, Yang, Cui, Yi and Que, 2025). In another study, Manoj and colleagues proposed a blockchain-based framework called griFLChain. Within this framework, they used deep learning models to predict crop yields in agriculture, recorded the models' metadata on the blockchain, and provided transparency and traceability through smart contracts (Manoj, Makkithaya and Narendra, 2025).

In another study, Jamshed et al. proposed a conceptual framework for a private blockchain, based on the Ethereum platform, for property transactions using proof of authority (Jamshed, Waheed, Iqbal, Faheem, Ashraf and Mansoor, 2025). In their study, Zhang and colleagues proposed a real estate investment tokenisation platform based on blockchain technology to redefine investment and transaction methods in the real estate market. They aimed to simplify the investment process, lower participation barriers, and improve transparency by converting real estate assets into digital tokens (Zhang, Ci, Wu and Wiwatanapataphee, 2025).

## BLOCKCHAIN

Blockchain, first recognised for its use in 2008 with Bitcoin (Nakamoto, 2008), a digital currency proposed by Satoshi Nakamoto, is a decentralised data management technology with a distributed consensus mechanism that removes the need for third-party validation of peer-to-peer network transactions. Data is stored encrypted within blocks to ensure confidentiality and security. Blocks are lists of records linked together in a chain data structure.

Participating members have full authority to monitor all transactions in the blockchain network in a peer-to-peer (P2P) manner (Da and Viriyasitavat, 2019). Blockchain technology spans several disciplines, including distributed computing, software development, and cryptography. The integration of these disciplines creates a secure infrastructure for digital assets (Berg, Davidson and Potts, 2019). Each new block in a blockchain must contain the hash of the previous block. When a new block is created, its header is hashed, and the resulting hash is checked against the required character set. If the block's hash value meets the required character set, the block is accepted and added to the chain. In this way, a chain structure of interrelated blocks is formed. A blockchain is categorised into three types based on its use: open, private, and hybrid (Casino, Dasaklis and Patsakis, 2019). In an open blockchain, anyone can freely join or leave the network and perform block creation and verification operations. In a private blockchain, these operations are restricted to a select group. Participation in the network may be public or restricted in various ways. In a hybrid blockchain, also known as a consortium blockchain, each node belongs to an organisation. Joining or leaving the network requires authorisation. Block creation and verification are managed by a single group of organisations. All organisations form a coalition to maintain operations together (Zhang, Zhong, Wang, Chao, and Wang, 2020).

## THE GATECHAIN APPLICATION

In this section, we present the GateChain application. GateChain is a permissioned blockchain-based application developed to ensure reliable, traceable, and transparent entry and exit procedures within a country. Only authorised institutional personnel can access and record data through the application. The application offers the following solutions:

1. It speeds up verification in passport control processes.
2. It prevents record manipulation.
3. It enables reliable data sharing between different institutions.

GateChain is an innovative solution that supports digital transformation in national security applications by integrating blockchain capabilities into traditional border-crossing systems. All transactions at border crossings are recorded on the blockchain as new blocks. A person's entry transaction is stored in one block, and their exit transaction in another. It is not possible to update an existing block on the blockchain during the exit process, as doing so would incur extremely high computational costs. This performance cost is also one of the most important factors ensuring the security of the blockchain.

**Figure 1**. Entry-exit registration screens

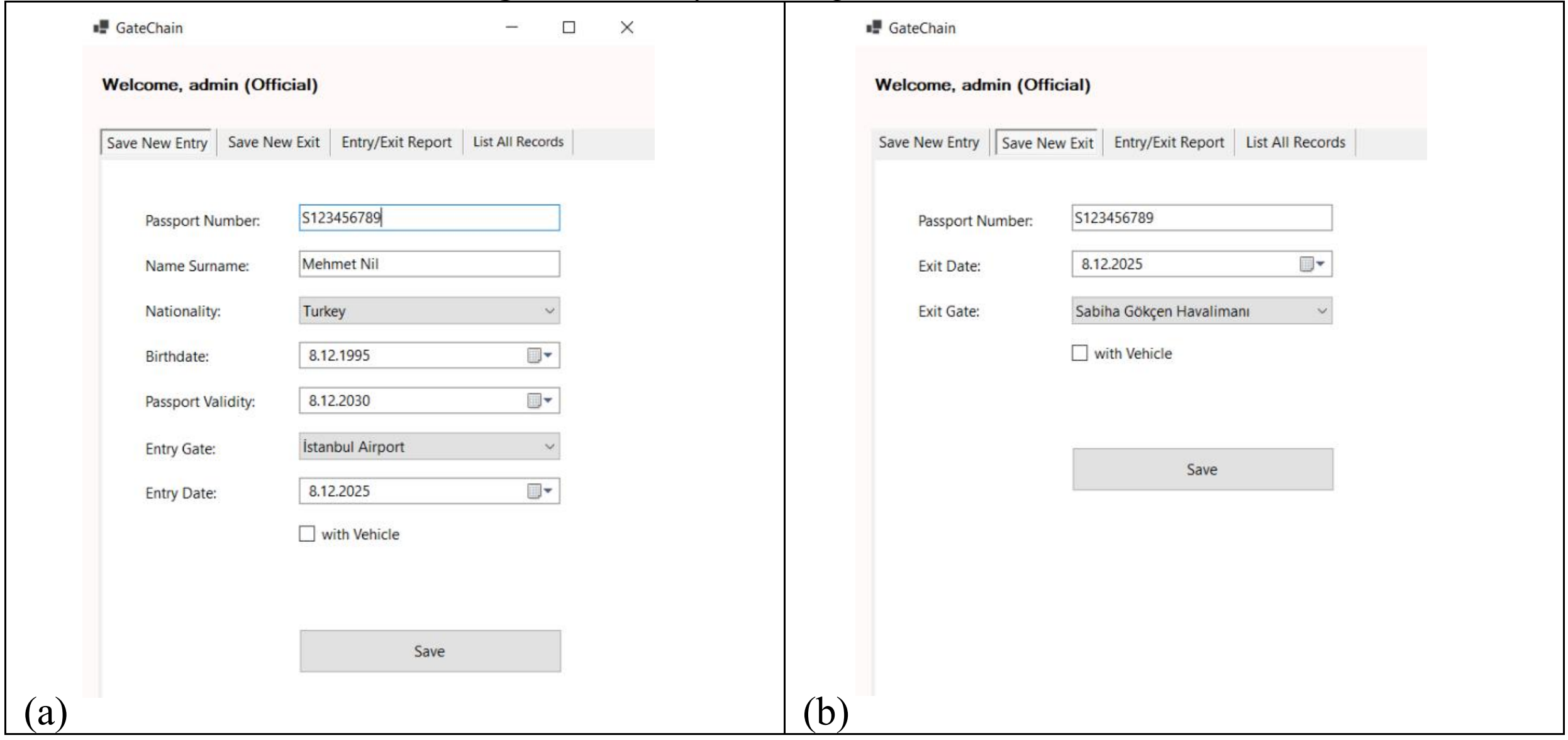

Figure 1a shows the application screen where personal information is recorded on the blockchain upon entry into a country. Figure 1b shows the application screen used when a person exits the country. Although a person's entry and exit transactions are stored in different blocks on the blockchain, they are displayed as a single record for listing and reporting purposes.

**Figure 2**. List All Records Screen

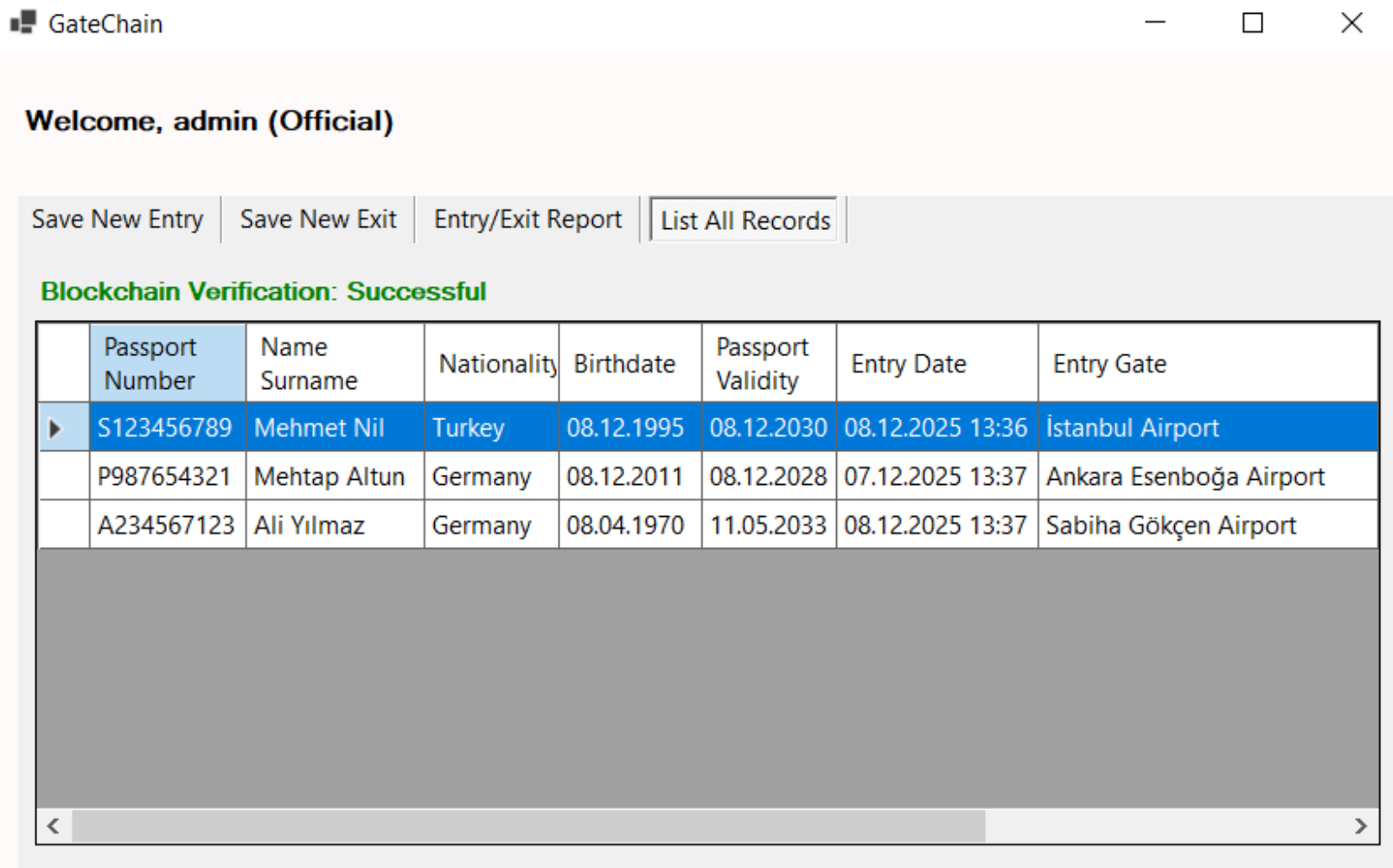

Figure 2 shows the screen listing all records in the blockchain. This screen also enables verification and validation of the blockchain. The application panel provides authorities with access to blockchain data, record queries, and statistical reporting functions. In a GateChain application, the number of validators on the network must be small and predetermined. Furthermore, at high-density border crossings, blockchain verification and block addition must be performed rapidly. Each block must also include proof of authority, consisting of the authority's public key and block signature. Therefore, Proof of Authority (PoA) is used as the consensus mechanism within the application.

A new block is added to the blockchain for each user transaction. Each block records the passport number, full name, date of birth, passport expiry date, entry or exit date and time, entry or exit point, and vehicle licence plate (if applicable). Each block also stores a summary of the previous block, the current summary, a timestamp, a nonce, an authority, and a signature. Unlike in PoW, the nonce does not contain a counter value; it indicates the type of block. The authority includes the authorised user's public key. When an authorised user creates a block, they sign it with their private key. This signature is added to the block as signature data. The authority and signature data verify the identity of the user who signed the block.

The following is an example of a block added to the blockchain for an entry operation.

{
"authority":"05ec42e9e101e45bf08b47e548eb6c8992fa8351c5c00d295556c0251cd5aa554d1
be4987bc"
"hash": "c770c3ffc392f83f0f6bb8c7ffe7312b8d8b74448b054247a3db8c7c87557c9d",
"index": 3, "nonce": "0x00", "timestamp": 1765221317.255807,
"previousHash":
"38d78af90f662cfcb0265b55fe0cacb031ddf5435256cdc165ff65e7152d684d",
"signature":"a199ce3160fe6323da5868707adcd24d055be9b6f101321971d5a5f5bbad57db1dc
",
"transactions": [{"Birthdate": "1995-08-12", "EntryDate": "2025-12-08 13:36", "EntryGate": "Istanbul Airport", "ExitDate": "", "ExitGate": "", "NameSurname": "cAPkxMkFeJY44XT4f/Kw==", "Plate":"",
"Nationality": "4fvM51g4JebJlDcjmbLg==", "PassportValidityDate": "2030-08-12", "PassportNumber": "mRZqQQe0BYjSC6OqjNWxw=="}],
"transactions_root":"34835a729927172e07e7eabb16edbb797d596aa52677cdbd04840674d8"}

Many pieces of personal data on the blockchain are encrypted using AES-256, a private-key encryption algorithm. For authorisation operations, ECDSA, a public-key encryption method, is used for signing and verification.

## RESEARCH AND FINDINGS

This section presents performance and security evaluations of GateChain.

### Performance Evaluation

The GateChain application was tested for encryption, signing, total time, and estimated transactions per second (TPS) on the blockchain. Results were obtained by creating 1,000 blocks.

**Table 1**. Adding Blocks to a Blockchain: Transaction Operation Times

| Block Index | Encryption Time | Sign Time | Total Time | Estimated TPS |
|---|---|---|---|---|
| 1 | 0.001582 | 0.001823 | 0.006538 | 5256.899364 |
| 10 | 0.000074 | 0.002733 | 0.007048 | 2406.449815 |
| 50 | 0.000049 | 0.000599 | 0.005278 | 8056.531799 |
| 100 | 0.000009 | 0.001553 | 0.006264 | 22638.16190 |
| 250 | 0.000039 | 0.001672 | 0.018700 | 13992.00577 |
| 500 | 0.000033 | 0.000498 | 0.025545 | 21041.82644 |
| 750 | 0.000035 | 0.003334 | 0.037920 | 20763.08476 |
| 1000 | 0.000037 | 0.000370 | 0.047199 | 22194.83003 |

Table 1 presents the computation times and estimated TPS for adding a block to a blockchain, using several example blocks.

**Figure 3**. Encryption and Signing Times

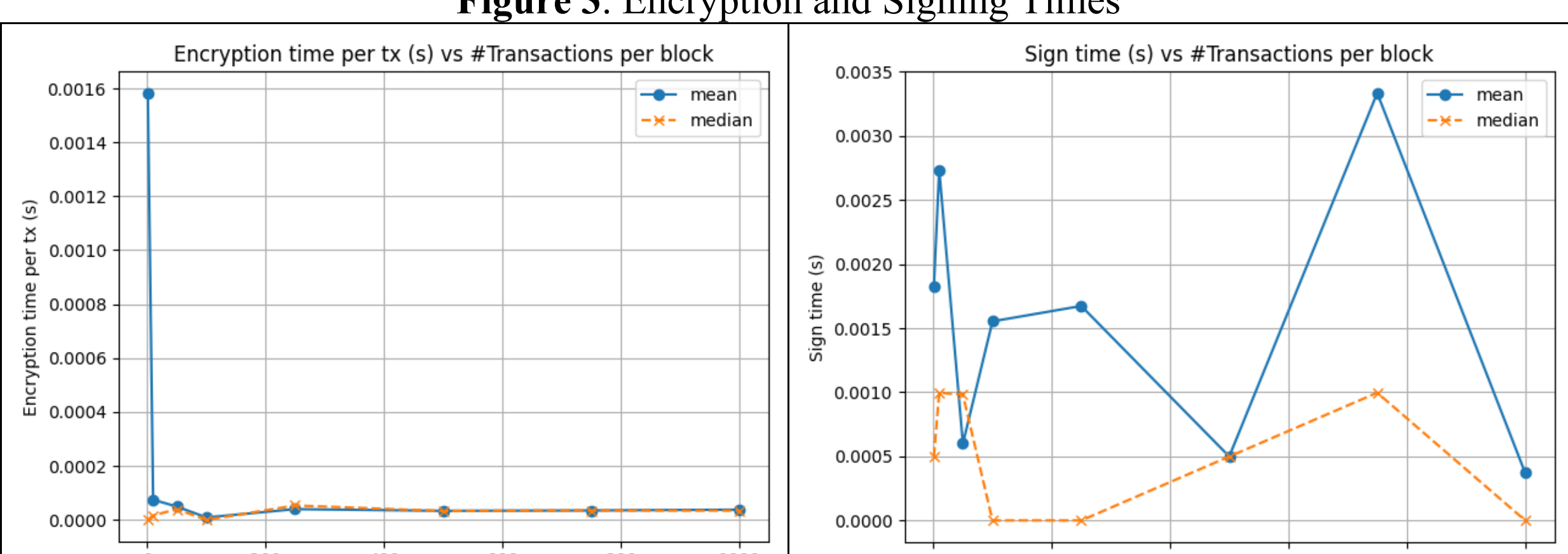

Figure 3a shows the time required to encrypt the generated blocks using AES-256. In real-time applications, where hundreds of thousands of entry and exit registrations occur daily, the cost of AES becomes critical. The average encryption time for 1,000 blocks is 0.000072 seconds per transaction. The average number of encryptions per second is 1/0.000072 = 13.888.

Figure 3b shows the time required for an authority to sign each block it creates. In PoA, block production times are primarily limited by the signing delay, which depends on the signing method and key length. In the GateChain implementation, the ECDSA method was chosen because it is faster than RSA and has smaller key and signature sizes at the same security level (Gupta, Gupta, Chang and Stebila, 2002, September; Hankerson, Vanstone and Menezes, 2004; Dimitoglou and Jim, 2023, July). The average signing time for 1,000 blocks is 0.001313 seconds per transaction. The number of signatures per second is 1/0.001313 = 761.6.

The signing process shown in both Table 1 and Figure 3b is performed when a block is created. Separately from the signing process, the block verification process requires the verification of authorised signatures for each block. Signature verification is also a significant performance metric affecting system cost. In practice, the average signature verification time for 1,000 blocks is 0.005664 seconds per transaction. The number of signature verifications per second is 1/0.005664 = 176.5. Signature verification is generally two to five times slower than the signing process.

**Figure 4**. Block Total Time and TPS Over Time

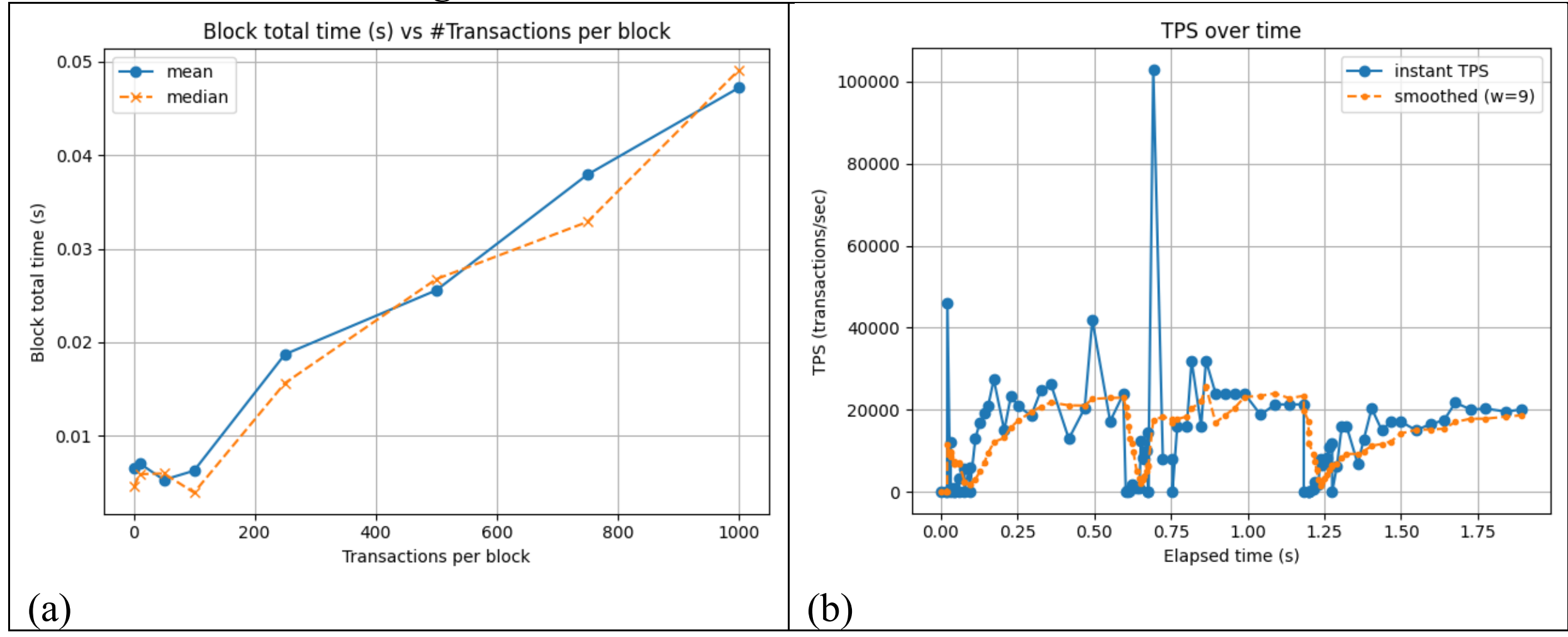

Figure 4a shows the total time required to create blocks from scratch and add them to the chain. As this time decreases, the TPS value increases. Figure 4b shows the approximate transaction capacity per second, calculated from the number of transactions per block and the block creation time. This value indicates how many entry and exit registrations can occur in the system per second.

These results demonstrate that GateChain can meet the requirements for real-time border-crossing verification.

**Security Evaluation**

Byzantine Fault Tolerance: As the application uses a PoA consensus mechanism, the system will continue to operate even if 33% of the authorised validators are malicious. For the blockchain to lose credibility, more than 50% of the registered validators would need to be malicious.

Authentication: The application eliminates the risk of unauthorised access by using a cryptographic authentication method defined for each authorised user. Furthermore, the actions that can be performed within the application are determined by user permissions.

Sybil Resilience: Thanks to the PoA consensus mechanism, attempts by unverified individuals to create fake authorisations have been neutralised.

Data Leak Prevention: As the records are stored in encrypted form, they cannot be read by unauthorised parties. Malicious actors would also need to obtain the key used in the encryption process.

Preventing Alterations: Each block on the blockchain contains the hash of the preceding block. Therefore, if an unauthorised change is made to a block's data, the previous hash of all subsequent blocks in the chain would need to be altered. This would incur a significant performance cost and is not feasible in blockchain structures with a large number of blocks. Furthermore, a malicious user would need to possess the AES encryption key used when the block was created to make any changes. Otherwise, the block data cannot be decrypted with the original key, and authorities would detect any alterations. Additionally, a block is signed by an authorised user when it is created, and the signature is recorded within the block as "signature" data. Any change within the block would require a change in the signature data, demonstrating that the transaction was not made by the authorised user who created the block.

## CONCLUSION

This study proposes GateChain, a blockchain-based application, for the secure, transparent, and auditable management of entry and exit information within a country. Given the weaknesses of traditional centralised structures, such as the risk of data manipulation and limited interoperability, GateChain's distributed, immutable ledger structure mitigates manipulation risks and enables secure data sharing between institutions. The application presented in this study demonstrates that a blockchain-based solution is applicable to border management processes.